\documentclass{ws-procs9x6-notrimmark}
\def\draftnote{~}

\usepackage{subfigure}

\newcommand{\vect}[1]{\vec{#1}}

\newcommand{\eqs}[1]{\begin{equation} \begin{split} #1\end{split} \end{equation} }

\newcommand{\ks}[1]{#1 \!\!\! \slash } 

\newcommand{\ga}{\gamma^5}

\newcommand{\ie}{{\it i.e.}}
\newcommand{\eg}{{\it e.g.}}
\newcommand{\etal}{{\it et al.}}

\newcommand{\cf}[1]{{Fig.~\ref{#1}}}

\newcommand{\beq}[1]{
\begin{equation}\label{#1}}
\newcommand{\eeq}{\end{equation}}
\newcommand{\bea}[1]{
\begin{eqnarray}\label{#1}}
\newcommand{\eea}{\end{eqnarray}}
\newcommand{\out}{\raise-3pt\hbox{\scriptsize    out}}

\begin{document}

\title{ Transition Distribution Amplitudes}

\author{J.P. Lansberg$^1$\footnote{lansberg@cpht.polytechnique.fr}, B. Pire$^1$ and L. Szymanowski$^{2,3}$}

\address{$^1$Centre de Physique Th\'eorique, \'Ecole Polytechnique, CNRS,
91128, Palaiseau, France\\
$^2$Fundamental Interactions in Physics and Astrophysics, Universit\'e de Li\`ege, 
Belgique\\
$^3$Soltan Institute for Nuclear Studies, Warsaw, Poland
}

\begin{abstract}
We give a brief overview of the theoretical status of the Transition Distribution Amplitudes and discuss
their experimental near future.

\end{abstract}

\keywords{Hadronic Physics, Exclusive Processes,  Factorisation}

\bodymatter

\section{Introduction}\label{introduction}

According to a now well-established framework \cite{Muller}, the Bjorken limit of 
near forward exclusive reactions with a hard probe allows factorisation of the leading twist
amplitudes into 
a perturbatively calculable sub-process at quark and gluon level on the one hand and 
hadronic matrix elements of light-cone non-local operators describing the transition 
 from the baryon target to a final baryon  on the other hand. Those matrix elements
can be expressed through hadron distribution amplitudes (DAs) and generalised parton 
distributions (GPDs).
Experimental data~\cite{Data2} from DESY
 and JLab are now confirming this framework, and they seem to show that its applicability 
 is quite precocious in terms of $Q^2$. The harmonic analysis \cite{asym} of spin 
 asymmetries which are particularly sensitive to the interference between the 
deeply-virtual-Compton scattering (DVCS) 
 and Bethe-Heitler processes is very relevant for that purpose. 

Investigations on GPDs are very important since they are new QCD objects
 which carry much information on the hadronic structure. A further generalisation of the GPD
  concept has been proposed \cite{pioneer} in cases where the initial and final states are 
different hadronic states.
  If those new hadronic objects are defined through a quark-antiquark operator (meson to meson or meson to photon transition), we call them {\it mesonic}
 transition distribution amplitudes (TDA)~\cite{TDApigamma},
  if they are defined through a three quark operator (baryon to meson or baryon to photon transition), 
we call them {\it baryonic} transition distribution 
amplitudes~\cite{TDApiproton}.
 
\begin{figure}[b!]
\centering{
\subfigure[$\gamma^\star \gamma \to A \pi$]{\includegraphics[height=5.1cm,clip=true]{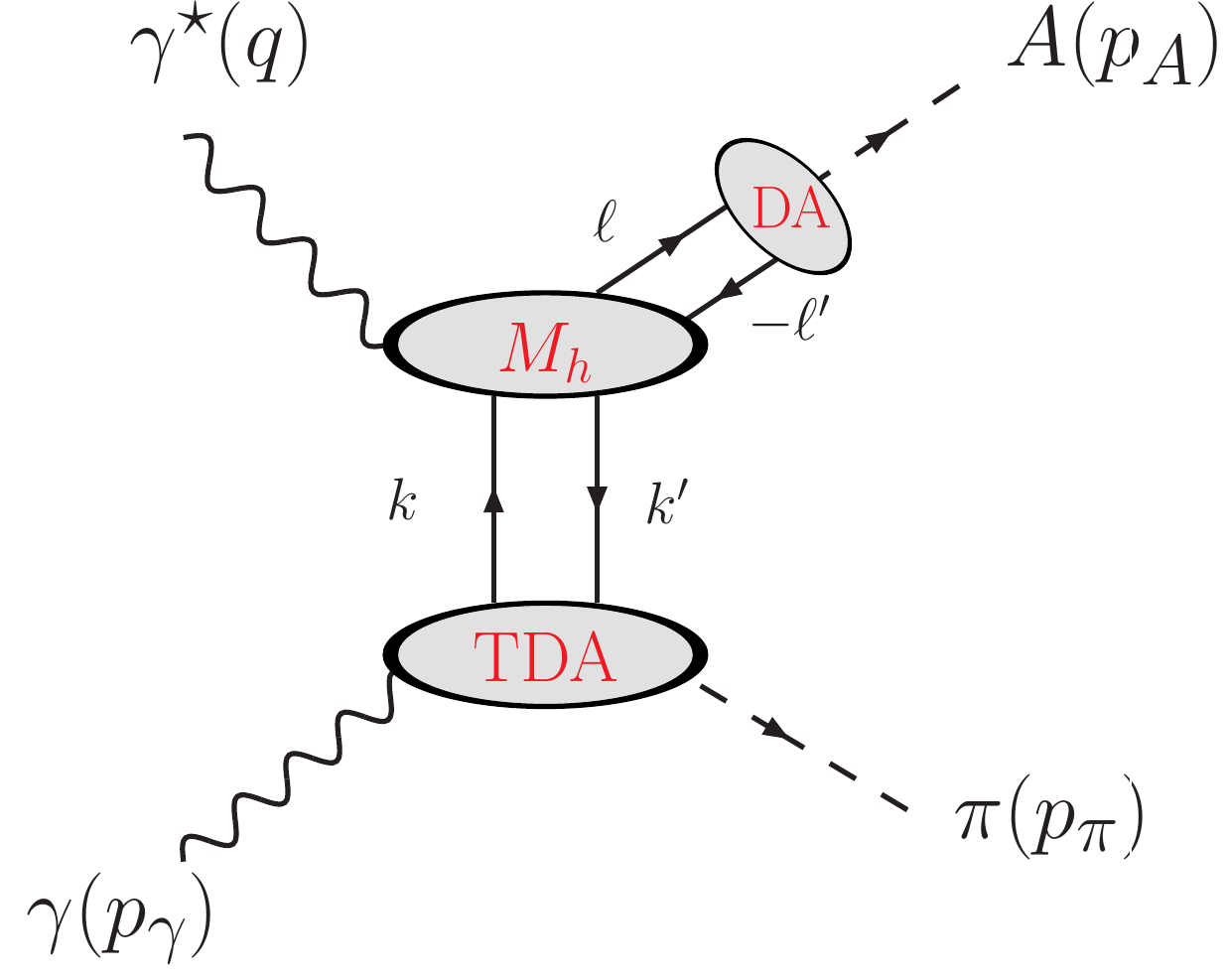}}
\subfigure[$\gamma^\star P\to P' \pi$]{\includegraphics[height=5.1cm,clip=true]{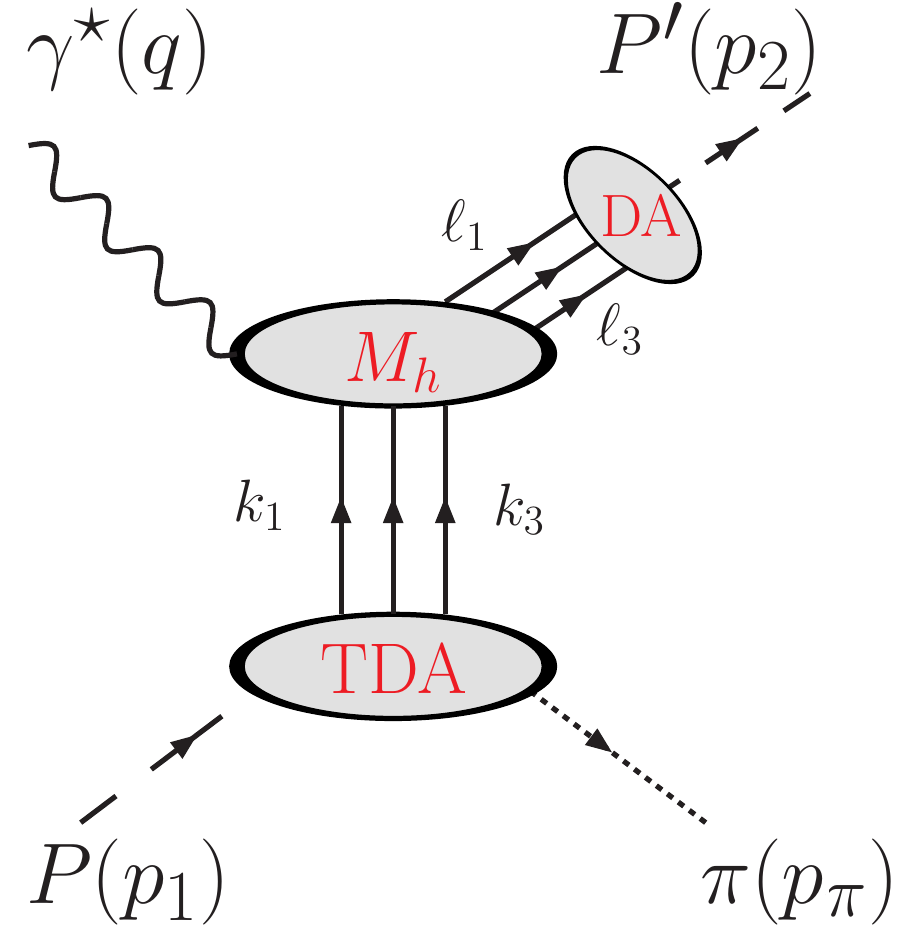}}
\subfigure[$\bar p p\to \gamma^\star \pi^0$]{\includegraphics[height=5.1cm,clip=true]{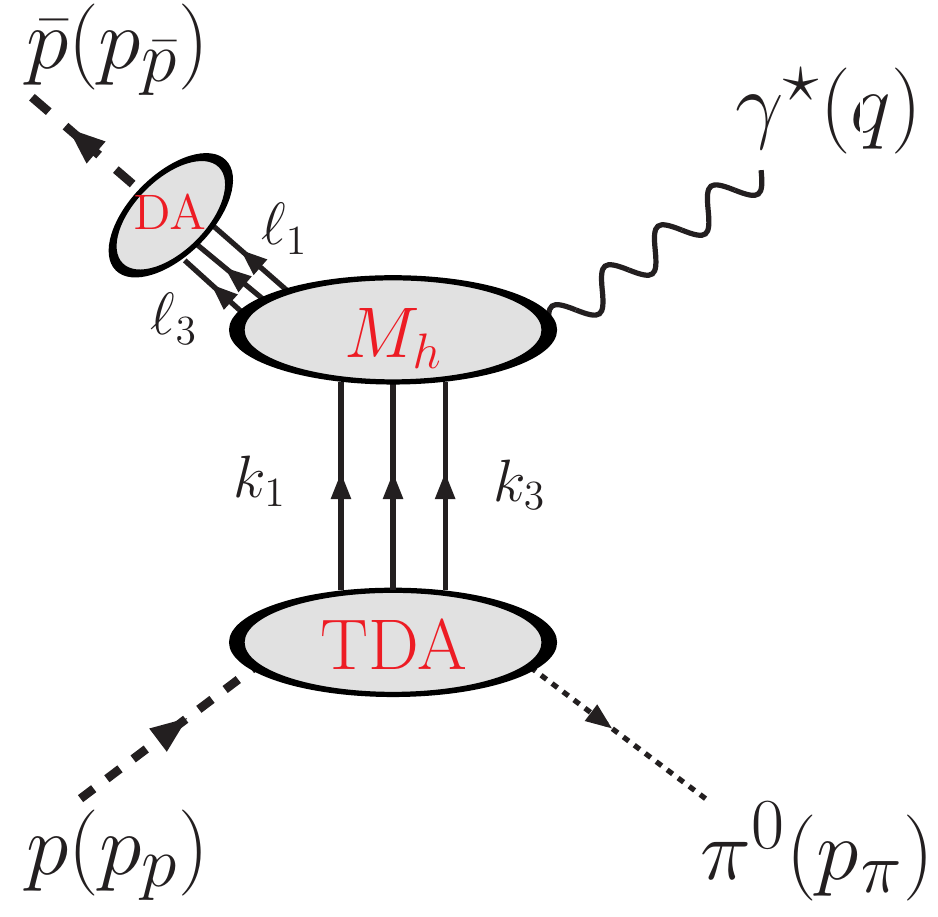}} 
}
\caption{(a) Illustration of the factorisation for $\gamma^\star \gamma \to A \pi$ at small transfer
momentum. (b) Idem for backward electroproduction of a pion. (c)  Idem for 
$\bar p p\to \gamma^\star \pi^0$.}
\label{fig:fact_illustr}
\end{figure}

Both appear in the description of hard exclusive processes, where a highly off-shell photon provides
a hard scale permitting, on the one hand, to treat perturbatively the interaction between the photon
and the quarks off the target, on the other hand to advocate the factorisation of
 the amplitude as a convolution
of a hard amplitude $M_h$ with the {\it universal} TDAs describing the non-perturbative
transition between two hadronic states and DAs describing the formation of another hadron.
This is illustrated in~\cf{fig:fact_illustr} for three cases.The first is the 
process $\gamma^\star \gamma \to  A \pi$ at small $t$, where $A$ is a meson. 
There appear the mesonic $\gamma \to \pi$ TDAs. The second is the 
backward electroproduction of a pion and the third is a crossed process 
$\bar p p\to \gamma^\star \pi^0$. In those two last processes, there appear 
the baryonic $p \to \pi^0$ TDAs.

\section{The mesonic TDAs}\label{mesonic}

\subsection{Mesonic TDAs vs. GPDs}

For definiteness, let us consider the $ \gamma  \to\pi^{-} $ TDAs. Those are defined
from the correlators\footnote{Working in light-like (or axial) gauge both for the gluons 
and the photons, we omitted to include the QCD and QED Wilson line between the quark fields.}
\footnote{If $p$ and $p'$ are respectively the momenta of the initial and final particles in the
TDAs, we have $(p'-p)=\Delta$, $t=\Delta^2$, $P=1/2(p+p')$ and  $2\xi =-\Delta^+/P^+$. We also have  
$\xi \approx {x_B}/{(2-x_B)}$.}
\begin{equation}
 \int \frac{d z^-}{2\pi} e^{ix P^+ z^-} 
\langle     \pi^-(p_{\pi^-})| \bar{d}(-\frac{z}{2})(\Gamma) u(\frac{z}{2})|
\gamma(p_\gamma,\varepsilon) \rangle \Big|_{z^+=0,\,  z_T=0},
\end{equation}
where $\Gamma$ is one of the Dirac forms, $\gamma^\mu$,$\gamma^\mu\ga$ or
$\sigma^{\mu\nu}$. The first two enable to respectively define the vector 
TDA $V^{\gamma\pi^-}(x,\xi,\Delta^2)$ and  the axial-vector 
one $A^{\gamma\pi^-}(x,\xi,\Delta^2)$, going along with their 
respective Lorentz structures~\cite{TDApigamma}.
The latter Dirac form, $\sigma^{\mu\nu}$, is related to two chiral-odd tensorial 
TDAs $T_1^{\gamma\pi^-}(x,\xi,\Delta^2)$ and $T_2^{\gamma\pi^-}(x,\xi,\Delta^2)$.

The mesonic TDAs have much in common with meson GPDs.
They are defined from matrix elements of the same quark - antiquark
operator and thus obey the same QCD evolution equations.   Sum rules  may  be derived  for  the  photon to meson
TDAs.  Since the local matrix elements  appear in radiative weak decays, we can relate the TDAs 
to the vector and axial 
form factors $F_V$ and $F_A$ in the $\pi^\pm$ case and 
to $F_{\pi^0\gamma^\star\gamma}$ in the $\pi^0$ case. Those form factors are 
well measured~\cite{Yao:2006px}.

 Moreover, the TDAs satisfy similar polynomiality conditions and they may be
 constructed from a spectral decomposition, in analogy with the construction of GPDs through  double 
 distributions~\cite{rad}. 
 The $x$ and $\xi$ dependence of the 
TDAs  is  then given as
\beq{DD}
\int_{-1}^{1} d\beta  \int_{-1+|\beta|}^{1-|\beta|}   d\alpha \; 
\delta(x-\beta -\xi \alpha)f(\beta,\alpha).
\nonumber
\eeq
In the GPD case, $f(\beta,\alpha)= q(\beta)h(\beta,\alpha)$ with $q(\beta)$  
 the forward quark distribution and $h(\beta,\alpha)$ a profile function.
Note however that TDAs possess different properties than GPDs with respect to time reversal
 since initial and final states are different. A consequence of this
is the appearance of odd-powers of $\xi$ in their moments in $x$.

  In the aforementioned case of $ \gamma  \to\pi^{-} $, 
the leading-twist decomposition of the matrix element differs 
from the $\pi^{-} $ GPD one mainly because of the presence of the photon polarisation vectors 
$\varepsilon(p_\gamma)$ -- a similar situation would occur for the $\pi\to\rho$ transition. This enables to build 4 independent leading-twist Lorentz structures, and thus to define 4 TDAs, whereas
there are only 2 leading-twist pion GPDs.

\subsection{A perturbative limit}
An interesting perturbative limit of a mesonic TDA may be derived \cite{Pire:2006ik} by considering the Born order amplitude
 ${\cal A}$ for  the process 
\begin{equation} \label{proces} 
\gamma^\star_L(q')\;\gamma^\star_L(q) \to \rho^0_L (k')\;\rho^0_L(k). 
\nonumber
\end{equation}
 in the region where  $Q'^2 \gg Q^2$ and  in the forward kinematics ($t=t_{min}$).

One can see that the amplitude factorises and the $\gamma^\star_L \to \rho_L$ 
vector $V(x,\xi,t)$
TDA\footnote{
Defining $n_1$ and $n_2$ such that $q=\frac{-Q^2}{(1+\xi)s} n_1+(1+\xi) n_2$ and $k=(1-\xi)n_2$ and
considering  the quark antiquark operator along the light-cone vector $n = 2n_1/s, s=2n_1.n_2$, 
$V(x,\xi,t)$ is defined as (omitting the QED Wilson line)
\eqs{\nonumber
\int\frac{dz^-}{2\pi}e^{ix(P.z)}\langle \rho^q_L(k)|
\bar q(-\frac{z}{2})\ks n
q(\frac{z}{2})|\gamma^\star(q)\rangle= \frac{e Q_q \epsilon_\nu(q)}{P^+Q^2}
[(1+\xi)n_2^\nu +\frac{Q^2n_1^\nu}{s(1+\xi)}]V(x,\xi,t)
.}
} is obtained at leading order as
\begin{equation}
\label{tda}
V(x,\xi,t_{min})= 3[\Theta\big(1\ge x \ge \xi) \Phi(\frac{x-\xi}{1-\xi}\big) -
\Theta(-\xi \ge x \ge -1)  \Phi\big(\frac{1+x}{1-\xi}\big)],
\nonumber
\end{equation}
where  $\Phi$ is the $\rho$ DA. This TDA describes the transition between the {\it least off-shell}
 photon ($Q^2$) and the longitudinally polarised $\rho$ to which it is collinear. Note that this perturbative 
expression vanishes in the ERBL region.

\subsection{Models}

We have already at our disposal several models for the vector and axial-vector TDAs describing
the pion to photon transition. In Refs~\refcite{Tiburzi,TDApigamma-appl}, those were modelled
via double-distributions and more recently in the spectral quark model~\cite{Broniowski:2007fs}. 
Other approaches~\cite{GPD_pion} used to construct pion GPDs could provide us 
with reasonable modelling of the mesonic TDAs. Let us cite for instance the Nambu-Jona Lasinio 
Model~\cite{Noguera}. Finally, lattice QCD, which has been recently applied to extract moments
of pion GPDs~\cite{Brommel:2007xd}, could also be applied to the TDA case.

\subsection{Experimental possibilities}
The introduction of the $\gamma \to $meson TDAs completes the kinematical domain of 
understanding of  the reactions  $\gamma \gamma^\star \to M_{1} M_{2}$ in 
the framework of QCD factorisation, supplementing  the near threshold kinematical 
domain described by generalised distribution amplitudes (GDAs)\cite{GDA}
and the fixed (or large) angle domain historically described by the Brodsky-Lepage factorisation \cite{BL}.
Data have been collected at LEP and CLEO on these reactions, mostly in the GDA domain 
for $\rho \rho$ final states, with some phenomenological success \cite{Anikin}. More 
data are obviously needed and are eagerly waited for in the TDA region, and much hope comes
from the high luminosity electron colliders. This requires in general the exclusive detection
of two mesons, when tagging simultaneously one outgoing electron.

Another way to access the mesonic TDAs is to study DVCS on virtual pion target, which 
may be studied\cite{amrath} at  Hermes and JLab in the reaction
$\gamma^\star p \to \gamma \pi^+ n$, when the transition $p\to n$ is dominated by the pion pole and the 
$\pi^+ $ flies in the direction of the $\gamma^\star$ in the $\gamma \pi^+$ CMS. 

\section{The baryonic TDAs}\label{baryonic}
\subsection{Definitions}
Let us concentrate here on the $p\to \pi$ TDAs. 
The leading twist TDAs for the $p \to \pi^0$ transition are defined from the correlator :
\begin{equation}
  \langle     \pi^{0}(p_\pi)|\, 
\epsilon^{ijk}u^{i}_{\alpha}(z_1\,n) u^{j}_{\beta}(z_2\,n)d^{k}_{\gamma}(z_3\,n)
\,|p(p_p,s_p) \rangle 
\nonumber
\end{equation}
These TDAs are matrix elements of the same operator that appears in baryonic distribution amplitudes. The 
known evolution equations of this operator lead to derive evolution equations which have different 
forms in different regions ; one defines one ERBL and two DGLAP regions
much in the same spirit as in the GPD case, so that the evolution equations in momentum space depend on 
the signs of the quark momentum fractions $x_{i}$.

As for DAs, an asymptotic solution for this evolution equation exists but the phenomenological study of
electromagnetic form factors leads us to strongly doubt that it is of any phenomenological relevance. 
In some sense, this is not a surprise since the corresponding asymptotic solution ($\delta(x)$) for parton 
distribution functions is far from a realistic description of DIS data. We thus do not propose to take an 
asymptotic TDA as a realistic input for phenomenology.

On the other hand, there exists an interesting soft limit \cite{Lansberg:2007ec} when the emerging pion momentum is small, which allows to relate proton $\to$
pion TDAs to proton DAs. The well-known soft pion theorems indeed allow to write:
\begin{equation}
\langle \pi^a(p_\pi)  |{\cal O}| P(p_1,s_1)\rangle \to -\frac{i}{f_\pi} \langle 0  | [ Q^a_5,  {\cal O}]  | P(p,s) \rangle \\ \nonumber
 \nonumber
 \label{eq:soft-theorem}
\end{equation}
when  $\xi \to 1$ ($E_\pi\to 0$) ; the neglected  nucleon pole term, which does not contribute at 
threshold but is
likely to be important for $\xi$ significantly different from 1, may also be taken into account. 
One then gets  relations
between the nucleon DAs~\cite{CZ} $A^p$, $V^p$ and $T^p$ on the one hand and the $p\to \pi$ TDAs $V^{p\pi^0}_{1}$, 
 $A^{p\pi^0}_{1}$ and $T^{p\pi^0}_{1}$ on the other hand :
\begin{eqnarray}
&&V^{p\pi^0}_1(x_1,x_2,x_3,1,M^2) =\frac{1}{4}  V^p  \Big(\frac{x_1}{2},\frac{x_2}{2},\frac{x_3}{2} \Big), 
\nonumber \\
&&A^{p\pi^0}_1(x_1,x_2,x_3,1,M^2) =\frac{1}{4} A^p  \Big(\frac{x_1}{2},\frac{x_2}{2},\frac{x_3}{2} \Big), 
 \nonumber\\
&&T^{p\pi^0}_1(x_1,x_2,x_3,1,M^2) =\frac{3}{4}T^p  \Big(\frac{x_1}{2},\frac{x_2}{2},\frac{x_3}{2} \Big). 
 \nonumber  
\end{eqnarray}
To conclude, let us mention that the proton to photon TDAs, entering the description of
backward DVCS, have been defined in Ref.~\refcite{Lansberg:2006uh}.

\subsection{Experimental situation}

As we mentioned above, $p\to \pi$ baryonic TDAs appear in
the description of backward electroproduction of a pion on a proton target. In terms of
angle, in the $\gamma^\star p$ center of momentum (CM) frame, the angle between
the $\gamma^\star$ and the pion, $\theta^\star_\pi$, is close to 180$^\circ$.
We then have $|u|\ll s$ and $t\simeq -(s+Q^2)$, in contrast to the fixed angle regime 
$u\simeq t\simeq -(s+Q^2)/2$ ($\theta^\star_\pi\simeq 90^\circ$) and the forward (GPD) one
 $|t|\ll s$ and $u\simeq -(s+Q^2)$ ($\theta^\star_\pi\simeq 0^\circ$).

The TDAs appear also in similar electroproduction processes such 
as $e p \to  e\;(p,\Delta^+) \; (\eta,\rho^0)$,  $e p \to  e\;(n,\Delta) \; (\pi^+,\rho^+)$, $e p \to  e\;\Delta^{++}\; (\pi^-,\rho^-)$. Those processes have already been analysed, at
backward angles, at JLab in the resonance region, \ie~$\sqrt{s_{\gamma^\star p}}=W<1.8$ GeV, in order to study the baryonic 
transition form factors in the $\pi$ channel~\cite{park} or in the $\eta$ 
channel~\cite{Armstrong:1998wg,Denizli:2007tq}. 
Data are being extracted in some channels above the 
resonance region. The number of events seems large enough to expect to get cross section measurements for 
$\Delta^2_T<1$ GeV$^2$, which is the region described in terms of TDAs. 
Hermes analysis\cite{hadjidakis} for forward electroproduction may  also be extended to larger values of  
$-t$.  It has to be noted though that present studies are limited to $Q^2$ of order a few GeV$^2$, 
which gives no guarantee to reach the TDA regime yet.
Higher-$Q^2$ data may be obtained at JLab-12 GeV and in muoproduction at Compass within the next few years.
Besides comparisons with forthcoming experimental data, one may also consider results from 
global Partial Wave Analysis (\eg~SAID\cite{Arndt:2006ym}).

Crossed reactions involving TDAs in proton-antiproton annihilation (GSI-FAIR\cite{Spiller:2006gj}), 
with time-like photons (\ie~di-leptons) can also  be studied
with other mesons 
than a pion, \eg~$\bar p p \to  \gamma^\star\; (\eta,\rho^0) $, or on a different target
than proton $\bar p N \to \gamma^\star\pi$. Finally, one may also consider 
associated $J/\psi$ production with a pion  $\bar p p \to  \psi\; \pi^0 $ or another 
meson $\bar p p \to  \psi\; (\eta,\rho^0) $,
which involve the {\it same} TDAs as with an off-shell photon or in backward electroproduction.
They will serve as  very strong tests of the universality of the TDAs in different processes.

\subsection{Models and applications}

The first application of baryonic TDAs was  centered on backward electroproduction of a 
pion~\cite{Lansberg:2007ec}. In that case, the hard contribution which consists in the 
scattering of the hard photon with three quarks is known at leading order.
Extrapolating the limiting value of the TDAs obtained from the soft pion theorems to the large-$\xi$ 
region, we obtained a first evaluation of the unpolarised cross section for 
backward electroproduction.This estimate, which is unfortunately reliable only in a 
restricted kinematical domain (large-$\xi$), shows an interesting sensitivity to the underlying 
model for the proton DA. This study will soon be extended to hard exclusive production of a $\gamma^\star \pi^0$ pair in $\bar p p $ annihilation at GSI-FAIR~\cite{ppgstarpi}.

In the near future, a  modelling inspired from the soft-pion limit including the nucleon pole term
will be applied to backward electroproduction at $\theta^\star_\pi\neq 180^\circ$. Application of the Pion-cloud Model~\cite{Pasquini:2006dv} to $p\to \pi$ transitions are also expected to be 
available soon\cite{Pasquini}. Specific non-perturbative approaches 
such as lattice studies, instanton-based models, chiral perturbation theory~\cite{Bernard:2006gx}, AdS-CFT correspondence~\cite{Brodsky:2003px}, can also provide us with realistic expression for the TDAs
to be checked experimentally and are therefore eagerly waited for.

\section*{Acknowledgments}
L.Sz. acknowledges the support by the Polish Grant 1 P03B 028 28.
L.Sz. is a Visiting Fellow of FNRS (Belgium).This  work  was partly supported by
the Joint Research
Activity "Generalised Parton Distributions" of the European I3 program
Hadronic Physics, contract RII3-CT-2004-506078.


\end{document}